  \documentclass[aps,prl,twocolumn,superscriptaddress,showpacs,floatfix,preprintnumbers,amsmath,amssymb]{revtex4}

\usepackage{epsfig}
\usepackage{rotating}
\usepackage{amsmath,amssymb}
\usepackage{relsize}
\RequirePackage{xspace}

\def\babar{\mbox{\slshape B\kern-0.1em{\smaller A}\kern-0.1em B\kern-0.1em{\smaller A\kern-0.2em R}}}

\def\to    {\ensuremath{\rightarrow}\xspace}
\def\invfb   {\ensuremath{\mbox{\,fb}^{-1}}\xspace}

\def\ccbar {\ensuremath{c\overline c}\xspace}
\def\ssbar {\ensuremath{s\overline s}\xspace}
\def\pep2{PEP-II}
\def\Bbar    {\kern 0.18em\overline{\kern -0.18em B}{}\xspace}
\def\BB      {\ensuremath{B\Bbar}\xspace} 
\def\BR      {{\ensuremath{\cal B}\xspace}}

\def\Omegapi      {\Omega_c^0\to\Omega^-\pi^+}
\def\OmegapipiZ   {\Omega_c^0\to\Omega^-\pi^+\pi^0}
\def\Omegapipipi  {\Omega_c^0\to\Omega^-\pi^+\pi^+\pi^-}
\def\XiKpipi      {\Omega_c^0\to\Xi^-K^-\pi^+\pi^+}
\def\FRatioOmegapipiZ {\frac{\BR(\OmegapipiZ)}{\BR(\Omegapi)}}
\def\FRatioOmegapipipi {\frac{\BR(\Omegapipipi)}{\BR(\Omegapi)}}
\def\FRatioXiKpipi {\frac{\BR(\XiKpipi)}{\BR(\Omegapi)}}

\newcommand{\BABARPubYear}    {07}
\newcommand{\BABARPubNumber}  {010}

\newcommand{\SLACPubNumber} {12398}
\newcommand{\LANLNumber} {0703030}

\begin{document}

\preprint{\babar-PUB-\BABARPubYear/\BABARPubNumber} 
\preprint{SLAC-PUB-\SLACPubNumber} 
\preprint{hep-ex/\LANLNumber}

\title{
{\large \bf Production and Decay of $\Omega_c^0$}
}

%
\author{B.~Aubert}
\author{M.~Bona}
\author{D.~Boutigny}
\author{Y.~Karyotakis}
\author{J.~P.~Lees}
\author{V.~Poireau}
\author{X.~Prudent}
\author{V.~Tisserand}
\author{A.~Zghiche}
\affiliation{Laboratoire de Physique des Particules, IN2P3/CNRS et Universit\'e de Savoie, F-74941 Annecy-Le-Vieux, France }
\author{J.~Garra~Tico}
\author{E.~Grauges}
\affiliation{Universitat de Barcelona, Facultat de Fisica, Departament ECM, E-08028 Barcelona, Spain }
% ...
\collaboration{The \babar\ Collaboration}
\noaffiliation

\date{\today}

\begin{abstract}
We present an analysis of inclusive $\Omega_c^0$ baryon production and decays
in 230.5~\invfb of data recorded with the \babar\ detector.
$\Omega_c^0$ baryons are reconstructed in four final states
($\Omega^- \pi^+$, $\Omega^- \pi^+ \pi^0$, $\Omega^- \pi^+ \pi^+ \pi^-$,
$\Xi^- K^- \pi^+ \pi^+$) and the corresponding ratios of branching fractions
are measured. We also measure the momentum spectrum
in the $e^+ e^-$ center-of-mass frame.
From the spectrum, we
observe $\Omega_c^0$ production from $B$ decays and in \ccbar events,
and extract the two rates of production.
\end{abstract}


\maketitle

The $\Omega_c^0$ ($css$) is the heaviest weakly-decaying
singly-charmed baryon. It has been observed
independently in several decay modes by different
experiments~\cite{bib:PDGbook} and in a variety of
production environments, including 
  $e^+ e^-$ colliders operating at the $\Upsilon(4S)$ resonance~\cite{bib:argus92,bib:cleo01,bib:cleo02},
  photoproduction~\cite{bib:focus93,bib:focus94,bib:focus03}, and
  hyperon beams~\cite{bib:wa89}.
So far, $B$ meson decays to $\Omega_c^0$ have not been observed.
Several different mechanisms could
contribute, principally weak decays of the following forms:
  $b \rightarrow c \bar{c} s$ (e.g., $B^- \rightarrow \Omega_c^0 \overline{\Xi}_c^-$);
  $b \rightarrow c \bar{u} s$ (e.g., $B^- \rightarrow \Omega_c^0 \overline{\Sigma}^-$); and
  $b \rightarrow c \bar{u} d$ (e.g., $B^- \rightarrow \Omega_c^0 \overline{\Xi}^0 \pi^-$).
Beyond the requirement to produce at least one $s\bar{s}$ pair during fragmentation,
we would expect these three types of decays to be further suppressed
  by the limited phase space, 
  by $|V_{us}|^2$, and
  by needing to produce a second $s\bar{s}$ pair, respectively.
Theoretical predictions for branching fractions of individual two-body
contributions vary from $\mathcal{O}(10^{-5})$ to
$\mathcal{O}(10^{-3})$~\cite{Chernyak:1990ag,Sheikholeslami:1991fa,Ball:1990fw}.

In this letter, we present a study of the $\Omega_c^0$ baryon,
reconstructed in four decay modes:
  $\Omega^- \pi^+$,
  $\Omega^- \pi^+ \pi^0$,
  $\Omega^- \pi^+ \pi^+ \pi^-$, and
  $\Xi^- K^- \pi^+ \pi^+$~\cite{footnote:cc}.
We measure the ratios of branching fractions for these modes,
normalizing to $\mathcal{B}(\Omega_c^0 \rightarrow \Omega^- \pi^+)$.
%
%
The previous most precise measurements of these ratios are from an
analysis of approximately 45 events from six $\Omega_c^0$ decay
modes~\cite{bib:cleo01}.
We then measure the spectrum of the $\Omega_c^0$ momentum in the
$e^+ e^-$ center-of-mass frame ($p^*$) and observe significant
production of $\Omega_c^0$ baryons in the decays of $B$ mesons.

The data for this analysis were recorded with the \babar\ detector at
the Stanford Linear Accelerator Center \pep2\ asymmetric-energy
$e^+e^-$ collider. The detector is described
in detail elsewhere~\cite{bib:babar}.
A total integrated luminosity of
230.5~\invfb is used, of which 208.9~\invfb were collected at the
$\Upsilon(4S)$ resonance (corresponding to 232~million
\BB pairs) and 21.6~\invfb were collected 40 MeV below
the \BB production threshold.

Simulated events with the $\Omega_c^0$ decaying into the
relevant final states are generated for the processes
  $e^+e^- \rightarrow \ccbar \rightarrow \Omega_c^0 X$ and
  $e^+e^- \rightarrow \Upsilon(4S) \rightarrow \BB \rightarrow \Omega_c^0 X$,
where $X$ represents the rest of the event. The \textsc{pythia} simulation
package~\cite{bib:pythia}
is used for the
$\ccbar$ fragmentation and for $B$ decays to $\Omega_c^0$, and
the \textsc{geant4}~\cite{bib:geant4} package is used to simulate
the detector response.
To investigate possible background contributions,
additional samples of generic Monte Carlo (MC) events are used,
equivalent to 990~\invfb for $\Upsilon(4S)$ events
($e^+ e^- \rightarrow \Upsilon(4S) \rightarrow \BB$), plus 320\invfb
for \ccbar continuum events ($e^+ e^- \rightarrow \ccbar$) and
340~\invfb for light quark continuum events
($e^+ e^- \rightarrow q\bar{q}$, $q = {u,d,s}$).

The reconstruction of an $\Omega_c^0$ candidate begins by identifying a
proton, combining it with an oppositely charged track interpreted
as a $\pi^-$, and fitting the tracks to a common vertex to form a
$\Lambda$ candidate. The $\Lambda$ is then combined with a negatively charged
track interpreted as a $K^-$ ($\pi^-$) and fit to a common vertex
to form an $\Omega^-$ ($\Xi^-$)
candidate. For each intermediate hyperon ($\Lambda$, $\Xi^-$, $\Omega^-$)
we require the invariant mass to be within 4.5~MeV$/c^2$ of its nominal
value (corresponding to approximately 4, 3, and 3 times the detector
resolution, respectively).
We form $\pi^0$ candidates from pairs of photons in the
electromagnetic calorimeter, requiring the energy of each photon to
be above 80~MeV and the combined energy to be above 200~MeV. We
require the invariant mass of the $\pi^0$ candidate, computed at
the event primary vertex, to be in the range 120--150~MeV$/c^2$.

Each $\Omega^-$ ($\Xi^-$) candidate that passes the requirements is then
combined with one or three additional tracks that are identified as
pions or kaons as appropriate. For the
$\Omega^- \pi^+ \pi^0$ final state, we also combine the hyperon
and $\pi^+$ with a $\pi^0$.
The $\Omega_c^0$ candidate daughters are refit to a common vertex
with their masses constrained to the nominal values. From this
fit we extract the decay vertices and associated uncertainties
of the $\Omega_c^0$ and the intermediate hyperons,
the four-momenta of the particles, and the
$\Omega_c^0$ candidate mass. For each intermediate hyperon
we require a positive scalar product of the momentum vector
in the laboratory frame and the displacement vector from its
production vertex to its decay vertex.

To further suppress the background, we compute the likelihood ratio
  $\mathcal L = {\prod_{i} p^{\mathrm{S}}_{i}(x_i)} / \left[ 
  {\prod_{i} p^{\mathrm{S}}_{i}(x_i) + \prod_{i} p^{\mathrm{B}}_{i}(x_i)} \right]$
for each $\Omega_c^0$ candidate,
where the index $i$ refers to the likelihood variables $x_i$,
and $p_{i}(x_i)$ are the probability density functions
for signal~(S) and background~(B). 
For a given $\Omega_c^0$ candidate ${\cal L}$ has a
value between 0 and 1. 
The likelihood variables $x_i$ are
  the logarithm of the $\Omega^-$ or $\Xi^-$ decay length significance,
    which is defined as the distance between the production and
    decay vertices divided by the uncertainty on that distance;
  the momentum of the $\Omega^-$ or $\Xi^-$ in the $e^+ e^-$ rest frame;
  the total momentum of the mesons recoiling against
    the $\Omega^-$ or $\Xi^-$ in the $e^+ e^-$ rest frame; 
  and, for the $\Omega^- \pi^+ \pi^0$ mode, the $\pi^0$ momentum in the
    laboratory frame.
These variables (particularly the decay length significance) cover
the expected range effectively with a limited number of bins.
The distributions of these variables for the signal hypothesis
are derived from signal MC simulations, 
and for the background hypothesis from generic MC events
in which contributions from real $\Omega_c^0$ are excluded.
Separate distributions are used for each final state when measuring ratios
of branching fractions, and for each momentum range when measuring the
momentum spectrum.

\begin{figure}
  \begin{center}
    \epsfig{file=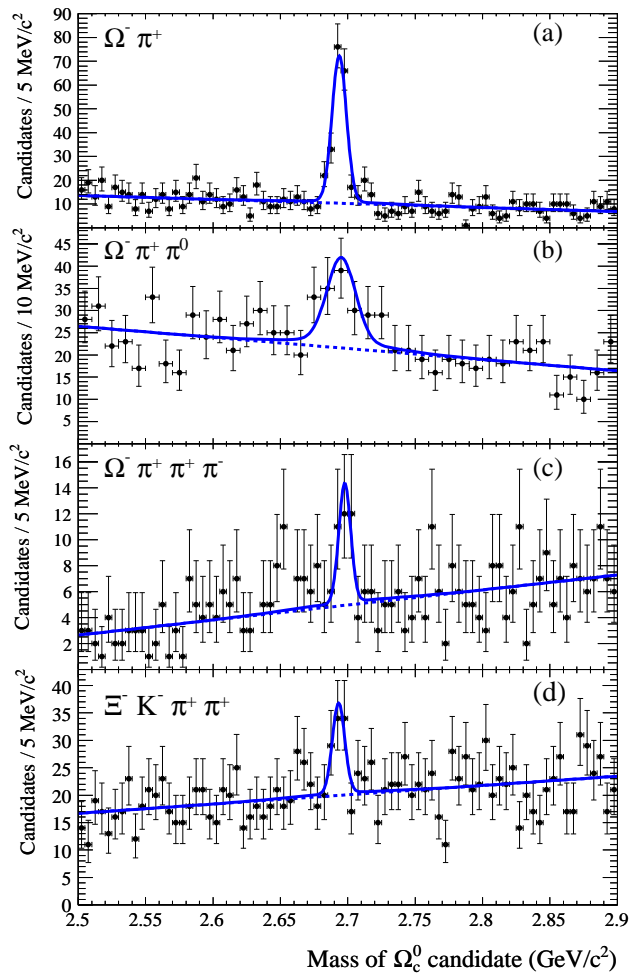, width=1.0\columnwidth}
  \end{center}
  \caption{
    The invariant mass spectra for candidates
    passing the selection criteria.
    The
    data are fit with a double Gaussian lineshape on a linear
    background.
  }
  \label{fig:ccdata}
\end{figure}

To measure the ratios of branching fractions, we require that 
$p^* > 2.4$~GeV$/c$ in order to suppress
combinatoric background. Since the
kinematic limit for $\Omega_c^0$ produced in $B$ decays
at \babar\ is $p^*_{\mathrm{max}} = 2.02$~GeV$/c$, only $\Omega_c^0$
produced in the \ccbar continuum are retained. 
We also require that the value of $\mathcal{L}$ for each
candidate is greater than a threshold $\mathcal{L}_0$, 
chosen to maximize the expected signal significance for a given
final state based on simulated events. 
We perform an unbinned maximum
likelihood fit to the mass distributions
shown in Fig.~\ref{fig:ccdata}.
The signal lineshape is parameterized as
the sum of two Gaussian functions with a common mean; the background is
parameterized as a first-order polynomial. 
In the fits to the data, the signal yield is a free parameter;
the widths and relative amplitudes of the two Gaussian functions are fixed
to values determined from a fit to simulated signal events.
The mean mass is also a free parameter, except for
the $\Xi^- K^- \pi^+ \pi^+$ final state where we fix it to the
central value obtained in $\Omega_c^0 \rightarrow \Omega^- \pi^+$
in order to ensure proper fit convergence.
The masses are found to be consistent with one another and with the
current world average~\cite{bib:PDGbook} within uncertainties.

The numbers of signal events are
  $177 \pm 16$, 
  $64 \pm 15$, 
  $25 \pm 8$, and
  $45 \pm 12$
(statistical uncertainties only) for the final states
  $\Omega^- \pi^+$,
  $\Omega^- \pi^+ \pi^0$,
  $\Omega^- \pi^+ \pi^+ \pi^-$, and
  $\Xi^- K^- \pi^+ \pi^+$, respectively.
These correspond to statistical significances of
  18, $5.1$, $4.2$, and $4.3$ standard deviations,
respectively, where the significance is defined as
$\sqrt{2 \Delta \ell}$ and $\Delta \ell$ is the change
in the logarithm of the likelihood between the
fits with and without an $\Omega_c^0$ signal component.
The fitted yields are then corrected for efficiency,
which is defined as the fraction of
simulated signal events, generated in the appropriate
$p^*$ range, that
are reconstructed and pass all selection criteria.
Including the loss of efficiency due to the $\Lambda$ and $\Omega^-$
branching fractions, we obtain efficiencies of
  $(8.6 \pm 0.6)\%$, 
  $(2.5 \pm 0.3)\%$, 
  $(4.3 \pm 0.4)\%$, and
  $(4.7 \pm 0.5)\%$
for the four final states,
where the uncertainties include systematic effects and are partially
correlated. The systematic uncertainties on, and corrections to, the
ratios of branching fractions are listed in 
Table~\ref{tab:ccsys} and discussed further later.
We measure the ratios to be
\begin{align*}
\FRatioOmegapipiZ  & = 1.27 \pm 0.31 \mathrm{(stat)} \pm 0.11 \mathrm{(syst)}, \\
\FRatioOmegapipipi & = 0.28 \pm 0.09 \mathrm{(stat)} \pm 0.01 \mathrm{(syst)}, \\
\FRatioXiKpipi     & = 0.46 \pm 0.13 \mathrm{(stat)} \pm 0.03 \mathrm{(syst)}.
\end{align*}

\begin{table}
  \caption{Systematic uncertainties on the ratios of branching fractions, where
$R_1 \equiv \mathcal{B}(\Omega_c^0 \rightarrow \Omega^- \pi^+ \pi^0)/\mathcal{B}(\Omega_c^0 \rightarrow \Omega^- \pi^+)$,
$R_2 \equiv \mathcal{B}(\Omega_c^0 \rightarrow \Omega^- \pi^+ \pi^+ \pi^-)/\mathcal{B}(\Omega_c^0 \rightarrow \Omega^- \pi^+)$, and
$R_3 \equiv \mathcal{B}(\Omega_c^0 \rightarrow \Xi^- K^- \pi^+ \pi^+)/\mathcal{B}(\Omega_c^0 \rightarrow \Omega^- \pi^+)$.
  }
  \begin{center}
    \begin{tabular}{lccc} 
    \multicolumn{1}{c}{Effect}                        & $R_1$ & $R_2$ & $R_3$ \\ \hline
    Finite MC sample size                              & 0.7\% & 0.7\% & 0.8\% \\
    Intermediate resonances in $\Omega_c^0$ decay      & 1.3\% & 2.6\% & 3.7\% \\
    Signal lineshape                                   & 1.0\% & 1.0\% & 1.0\% \\
    Dependence on the fit procedure                    & 1.5\% & 1.5\% & 1.5\% \\
    Hyperon branching fractions                        & ---   & ---   & 1.0\% \\
    Particle identification efficiency                 & 1.0\% & 1.0\% & 1.0\% \\
    Tracking efficiency                                & 0.0\% & 2.8\%\footnote{A relative correction of $+0.5\%$ applies to $R_2$ and $R_3$.}& 2.8\%\footnotemark[1] \\ 
    $p^*$ spectrum mismodeling                       & 1.5\% & 0.6\% & 3.5\% \\
    $\pi^0$ fitting and efficiency                     & 7.8\%\footnote{A relative correction of $+1.1\%$ applies to $R_1$.} & --- & --- \\
    \hline
    Total systematic uncertainty                       & 8.3\% & 4.4\% & 6.3\% 
  \end{tabular}    
  \end{center}
  \label{tab:ccsys}
\end{table}

\begin{figure}
  \begin{center}
    \epsfig{file=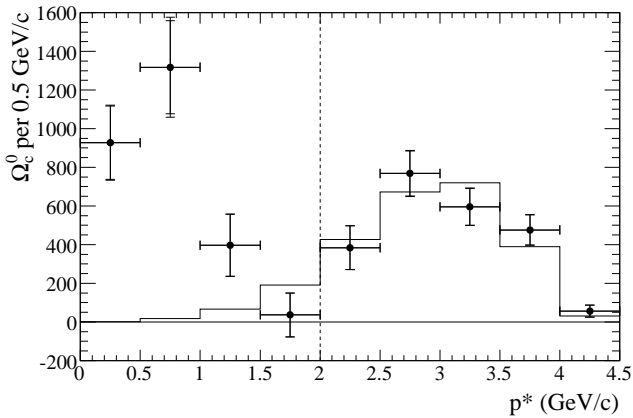, width=1.0\columnwidth}
  \end{center}
  \caption{
    The background-subtracted and efficiency-corrected
    $\Omega_c^0$ $p^*$ spectrum. The black points represent the
    data, with vertical error bars giving the sum in quadrature of
    statistical and uncorrelated systematic uncertainties. The
    solid histogram shows the Bowler fragmentation function, binned
    and fit to the data for $p^*>2$~GeV$/c$ (vertical, dashed line).
  }
  \label{fig:bbdata}
\end{figure}

We also measure the $p^*$ spectrum of $\Omega_c^0$ 
in order to study the production rates in both \ccbar
and \BB events. Only the $\Omega^- \pi^+$ final state is used.
The same reconstruction, optimization of selection criteria, and fitting procedures
described above
are applied, except that no requirement on $p^*$ is made. Instead,
the $\Omega_c^0$ candidates are divided into nine equal intervals
of $p^*$ covering the range 0.0--4.5~GeV$/c$. We again require
$\mathcal{L}>\mathcal{L}_0$ and compute
the efficiency in each $p^*$ interval as before with
simulated signal events. In the numerator of the efficiency we
count events with measured $p^*$ in the appropriate interval, and
in the denominator we count events with generated  $p^*$ in that
interval: this definition removes the slight broadening effect of
the detector momentum resolution. We also take into account a small
difference in efficiency between \ccbar and \BB events.
The efficiency-corrected
yield in each $p^*$ interval is shown in
Fig.~\ref{fig:bbdata}.

The systematic uncertainties
are divided into two
categories: normalization effects, which are treated as fully correlated
between all $p^*$ intervals, and shape effects, which are treated as
uncorrelated between different $p^*$ intervals. 
The normalization uncertainties are due to
  the mass resolution, which is determined from the MC and
       checked with studies of the control modes
       $\Xi_c^0 \rightarrow \Xi^- \pi^+$ and
       $\Xi_c^+ \rightarrow \Xi^- \pi^+ \pi^+$ (2.4\%);
  the $\Lambda$ and $\Omega^-$ branching fractions~\cite{bib:PDGbook}
       (1.3\%); and
  the tracking efficiency, which is corrected for data/MC
       discrepancies with control samples of $\tau$ decays (5.9\%).
The shape uncertainties are due to
  the limited size of MC samples ($<1\%$);
  dependence on the fit procedure (1.5\%);
  modeling of the $p^*$ spectrum, which
       can affect the weighted average efficiency within a $p^*$
       bin (0--6\%);
  the signal lineshape parameterization (1.0\%); and
  the particle identification efficiency (2.0\%).
When fitting fragmentation functions (see below), we consider only the
statistical and shape uncertainties, added in quadrature. When quoting
total yields and rates, we include the normalization uncertainties,
along with a relative correction of $+1.0\%$ due to a known 
data/MC discrepancy in tracking efficiency.

The double-peak structure seen in the $p^*$ spectrum is
due to two production mechanisms: the peak at lower
$p^*$ is due to $\Omega_c^0$ production in $B$ meson decays
and the peak at higher $p^*$ is due to $\Omega_c^0$ production
from the \ccbar continuum. This is consistent with the 
pattern observed in $\Lambda_c^+$ and $\Xi_c^0$ spectra measured for
$e^+ e^-$ annihilation at $\sqrt{s} = 10.6$~GeV~\cite{bib:belle_lambdac_pstar,bib:babar_lambdac_pstar,bib:xic_pstar}.
We fit the $p^*$ spectrum
with the Bowler fragmentation function~\cite{bib:bowler}
for $p^*>2$~GeV$/c$. 
We then extract the continuum yield
as the sum of the data points above 2~GeV$/c$ plus the integral of the
extrapolated function below 2~GeV$/c$. Similarly, the yield from
$B$ decays is the sum of the data points below 2~GeV$/c$ minus the
integral of the extrapolated function below 2~GeV$/c$.
Note that we do not fit a fragmentation function to the data below 2~GeV$/c$.
We
obtain yields of $2583 \pm 289$ and $2426 \pm 414$ for $\Omega_c^0$
produced in the continuum and in $B$ decays, respectively,
where the uncertainty includes all statistical and experimental effects.
An additional model uncertainty arises from the
extrapolation of the continuum tail for $p^* < 2$~GeV$/c$.
To estimate this, we repeat the $p^*$
spectrum fit and yield measurement with other fragmentation functions:
  Collins and Spiller (CS) \cite{bib:collins_spiller}, 
  two versions of the phenomenological model of Kartvelishvili {\it et al.} (KLP-M and KLP-B) \cite{bib:klpm,bib:klpb}
  and the Peterson model~\cite{bib:peterson}.
The CS and KLP-M fits are inconsistent with the data for $p^* > 2$~GeV$/c$.
The RMS of the yields from the three other fits is
240 events and is taken as the model uncertainty 
for the $B$ and continuum $\Omega_c^0$ yields.
Dividing the $\Omega_c^0$ yield in $B$ decays by the total number of $B$ mesons
in the data sample, we obtain the branching fraction product
$\mathcal{B}(B \to \Omega_c^0 X) \mathcal{B}(\Omega_c^0 \to \Omega^- \pi^+) = [5.2 \pm 0.9~\mathrm{(exp)} \pm 0.5~\mathrm{(model)}] \times 10^{-6}$,
where $X$ represents the rest of the $B$ meson decay products.
Dividing the $\Omega_c^0$ yield from the continuum by the
integrated luminosity and correcting for the small variation in
cross-section with $\sqrt{s}$, we obtain the cross-section product at
$\sqrt{s} = 10.58$~GeV:
$\sigma(e^+e^- \to \Omega_c^0 X) \mathcal{B}(\Omega_c^0 \to \Omega^- \pi^+) = \left[11.2 \pm 1.3~\mathrm{(exp)} \pm 1.0~\mathrm{(model)} \right] ~\mathrm{fb}$,
where $X$ represents the rest of the event.
As a cross-check, we also make model-independent
estimates of the yields from the continuum and from $B$ decays 
by subtracting the data below the $\Upsilon(4S)$ threshold.
Within large uncertainties, these are consistent
with the yields measured above.

It is thus clear that decays of $B$ mesons to $\Omega_c^0$ occur at a
significant rate. Assuming the absolute branching fraction
$\mathcal{B}(\Omega_c^0 \rightarrow \Omega^- \pi^+)$ 
$\sim 1\%$, we conclude that
  $\mathcal{B}(B \rightarrow \Omega_c^0 X) \sim \mathrm{few} \times 10^{-4}$. 
This is substantially lower than the inclusive $B$ meson branching
fractions to the charmed baryons $\Lambda_c^+$ and $\Xi_c$, which are
$\sim \mathrm{few} \times 10^{-2}$~\cite{bib:belle_lambdac_pstar,bib:babar_lambdac_pstar,bib:xic_pstar}.
One possible explanation for this is that both $\Lambda_c^+$ and $\Xi_c$
can be produced in a $b \rightarrow c \bar{c} s$ transition without creating
an $\ssbar$ pair from the vacuum, whereas at least one $\ssbar$ pair must be
created for $\Omega_c^0$ production. It is also possible that phase space suppression
in $B$ decays to baryons becomes significant when very close
to threshold.

In conclusion, we have studied the $\Omega_c^0$ baryon at
\babar\ through four hadronic decay modes, using 230.5~\invfb of data.
We measure the ratios of branching fractions for four modes,
significantly improving upon the previous values~\cite{bib:cleo01}.
We have also measured the $p^*$ spectrum and found comparable
production rates of $\Omega_c^0$ baryons from the continuum and from
$B$ meson decays. The inclusive $B$ branching fraction to $\Omega_c^0$
is found to be substantially lower than those to $\Xi_c^0$ and $\Lambda_c^+$
baryons, assuming the relevant baryon weak decay branching fractions
are of the same order of magnitude.

\begin{acknowledgments}
We are grateful for the excellent luminosity and machine conditions
provided by our \pep2\ colleagues, 
and for the substantial dedicated effort from
the computing organizations that support \babar.
The collaborating institutions wish to thank 
SLAC for its support and kind hospitality. 
This work is supported by
DOE
and NSF (USA),
NSERC (Canada),
IHEP (China),
CEA and
CNRS-IN2P3
(France),
BMBF and DFG
(Germany),
INFN (Italy),
FOM (The Netherlands),
NFR (Norway),
MIST (Russia),
MEC (Spain), and
PPARC (United Kingdom). 
Individuals have received support from the
Marie Curie EIF (European Union) and
the A.~P.~Sloan Foundation.
\end{acknowledgments}


\begin{thebibliography}{99}
\bibitem{footnote:cc}
  The use of charge conjugate modes is implied throughout.
\end{thebibliography}
\end{document}